\documentstyle[aps,prl,twocolumn,epsf,graphicx,amsmath,array]{revtex}

\begin{document}
\twocolumn[\hsize\textwidth\columnwidth\hsize\csname@twocolumnfalse%
\endcsname 
\title{Granular `glass' transition} 

\author{Leonardo E.~Silbert$^1$, Deniz Erta{\c s}$^2$, Gary
  S.~Grest$^1$, Thomas C.~Halsey$^2$, and Dov Levine$^3$}

\address{$^1$ Sandia National Laboratories, Albuquerque, New Mexico
  87185\\ $^2$ Corporate Strategic Research, ExxonMobil Research and
  Engineering, Annandale, New Jersey 08801\\ $^3$ Department of
  Physics, Technion, Haifa, 32000 Israel}

\maketitle

\begin{abstract}
  The transition from a flowing to a static state in a granular
  material is studied using large-scale, 3D particle simulations.
  Similar to glasses, this transition is manifested in the development
  of a plateau in the contact normal force distribution $P(f)$ at
  small forces, along with the splitting of the second peak in the
  pair correlation function $g(r)$, suggesting compaction and local
  ordering. The mechanical state changes from one dominated by plastic
  intergrain contacts in the flowing state to one dominated by elastic
  contacts in the static state. We define a staticity index that
  determines how close the system is to an isostatic state, and show
  that for our systems, the static state is not isostatic.
\end{abstract}
]

Granular materials are many-body systems which can exhibit properties
of both liquids and solids, often at the same time.  Features such as
jamming in silos and hoppers \cite{to1} and heterogeneous force
propagation \cite{mueth1} have motivated new proposals for
constitutive relations for granular materials \cite{cates2,cates9}, in
contrast to the established elasto-plastic theories
\cite{nedderman1,savage1}.  Moreover, the emerging field of ``jammed
systems'' \cite{liu1} seeks to understand whether analogies may be
drawn between diverse systems which have the common properties that
they are far from equilibrium, and unable to explore phase space.

O'Hern {\em et al.} \cite{nagel4} have recently shown that
similarities exist between the properties of {\it static} granular
packings and other amorphous systems; in particular, they compared
force distributions for simulated systems of soft-spheres undergoing a
glass transition with experimental data from static granular packings
\cite{mueth1,nagel5}. These studies suggest that it may be interesting
to consider the transition of granular materials from flowing to
static, and inquire about its similarities with the glass transition.
In this Letter we study the dynamic jamming transition of systems of
athermal grains through large-scale simulation.  In particular, we
consider the behavior of dense packings of granular particles flowing
down an inclined plane as the tilt angle $\theta$ is reduced through
the angle of repose $\theta_{r}$ at which flow ceases.

The chute flow geometry we employ relies on gravity to compactify the
system. Our earlier simulation studies in this geometry
\cite{deniz2,leo7} have proven their reliability by reproducing key
experimentally observed characteristics of dense granular flows
\cite{pouliquen1}, such as the existence of steady state flow over a
range of angles between $\theta_{r}$ and a maximum angle of flow
stability $\theta_{\rm max}$, as well as the dependence of the angle
of repose on pile height for shallow piles less than 20-30 grain
diameters in depth.

The transition from flowing to static states suggests that for systems
generated in this way, $\theta_{r}$ can be regarded as an analogue for
the glass transition temperature $T_{g}$, and we characterise the
transition by the following properties:
\begin{itemize}
\item $P(f)$, the probability distribution of normal forces, develops
  a plateau in the static state. 
\item The radial distribution function, $g(r)$ develops a split second
  peak, and growth of the first peak, indicating compaction and
  increased short-range ordering.
\item The average co-ordination number, $z_{c}$, jumps discontinuously.
  Moreover, the static state that results when motion ceases is not
  isostatic.
\end{itemize}

Furthermore, we observe a significant change in the number and nature
of interparticle contacts: although a finite fraction of contacts are
plastic (at Coulomb yield) for all flowing piles, almost none are
plastic when the pile becomes static \cite{footnote9}. The average
co-ordination number increases continuously towards four in the
flowing state and jumps to well above four for the static packing --
four being the theoretical minimal co-ordination number for a network
of frictional grains.

We carried out molecular dynamics simulations in 3D on a model system
of $N=8000$ mono-disperse, cohesionless, frictional spheres of
diameter $d$ and mass $m$. The system is spatially periodic in the
$xy$ (flow -- vorticity)-plane, and is constrained by a rough bottom
bed in the $z$-direction, with a free top surface. The static packing
height is about $40d$. Particles interact only on contact through a
Hertzian interaction law in the normal and tangential directions to
their lines of centers \cite{walton1,cundall1}. That is, contacting
particles $i$ and $j$ positioned at ${\bf r}_{i}$ and ${\bf r}_{j}$
experience a relative normal compression
\[\delta=|{\bf r}_{ij}-d|,\] 
where ${\bf r}_{ij}={\bf r}_{i}-{\bf r}_{j}$, which results in a force
\begin{equation}
{\bf F}_{ij}={\bf F}_{n}+{\bf F}_{t}.
\label{equation1}
\end{equation}
The normal and tangential contact forces are given by
\begin{equation}
{\bf F}_{n}=\sqrt{\delta/d}\left(k_{n}\delta 
{\bf n}_{ij}-\frac{m}{2}\gamma_{n}{\bf v}_{n}\right), 
\label{equation2}
\end{equation}
\begin{equation}
{\bf F}_{t}=\sqrt{\delta/d}\left(-k_{t}\Delta{\bf s}_{t}
-\frac{m}{2}\gamma_{t}{\bf v}_{t}\right),
\label{equation3}
\end{equation}
where ${\bf n}_{ij}={\bf r}_{ij}/r_{ij}$, with $r_{ij}=|{\bf
  r}_{ij}|$, ${\bf v}_{n}$ and ${\bf v}_{t}$ are the normal and
tangential components of the relative surface velocity, and $k_{n,t}$
and $\gamma_{n,t}$ are elastic and viscoelastic constants
respectively. $\Delta{\bf s}_{t}$ is the elastic tangential
displacement between spheres, obtained by integrating surface relative
velocities during elastic deformation of the contact \cite{footnote6}.
The magnitude of $\Delta{\bf s}_{t}$ is truncated as necessary to
satisfy a local Coulomb yield criterion $F_{t} \leq \mu F_{n}$, where
$F_t \equiv |{\bf F}_{t}|$ and $F_{n} \equiv |{\bf F}_{n}|$, and $\mu$
is the local particle friction coefficient.

In the steady state flow regime, changes in the particle friction
$\mu$, the damping coefficients $\gamma_{n,t}$, or particle hardness
$k_n$, do not change the qualitative nature of our
findings\cite{leo7}. For the present simulations we set
$k_{n}=2\cdot10^{5}mg/d$, $k_{t}=\frac{2}{7}k_{n}$,
$\gamma_{n}=50\sqrt{g/d}$, $\gamma_{t}=0.0$, and $\mu=0.5$
\cite{footnote5}. We choose a time-step $\delta t=10^{-4}\tau$, where
$\tau=\sqrt{d/g}$.

\begin{figure}[h]
\begin{center}
\includegraphics[width=6cm]{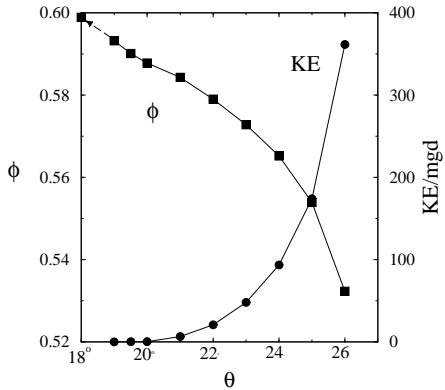}
\bigskip
\caption{\em Bulk packing fraction $\phi$ and bulk averaged kinetic energy
  (KE) per particle, normalised by $mgd$, as a function of tilt angle
  $\theta$. The dashed arrow indicates the value of $\phi$ when
  $\theta$ is reduced from $19^{\circ}$ to zero.}
\label{figure1}
\end{center}
\end{figure}

To initiate flow, we tilt the bed to a large angle ($\theta\approx
26^{\circ}$) long enough to remove any history effects of
construction.  We then lower the angle, in unit increments, to the
desired value in the range for which there is steady state flow,
$20^{\circ}\lesssim\theta\lesssim 26^{\circ}$ \cite{deniz2}. In this
range, the energy input from gravity is balanced by that dissipated in
collisions. We can approach the flowing angles from above and below
and always obtain the same results.

Lowering $\theta$, the kinetic energy of the system decreases, however
flow continues down to $\theta_{r} (\simeq19.4^{\circ}$ for the set of
parameters used here\cite{footgeneral}). As $\theta$ is reduced, the
volume fraction increases, as shown in Fig.~\ref{figure1}. We
generated further static packings by taking the stopped state at
$\theta=19^{\circ}$ and reducing the tilt to zero. The volume fraction
of the static packing for $\theta=0.0$, $\phi=0.599\pm0.005$, is less
than the random close packing value $\phi_{rcp}\simeq 0.64$.

In order to investigate the distribution of contact forces in flowing
and static states, we computed the probability density function (PDF)
$P(f)$, where $f\equiv F_{n}/{\overline F}_{n(z)}$ is the ratio of the
normal contact force magnitude $F_{n}$ to its average value at the
depth $z$ of each contact. Since our system is spatially periodic in
the horizontal plane, with no side-walls, the pressure does not
saturate with depth, requiring separate averaging at each value of
$z$. The data were averaged over 1000 configurations over a period of
$200\tau$ in the steady state at each flowing angle and 5
configurations for the static states.

\begin{figure}[h]
\begin{center}
  \includegraphics[width=6.5cm]{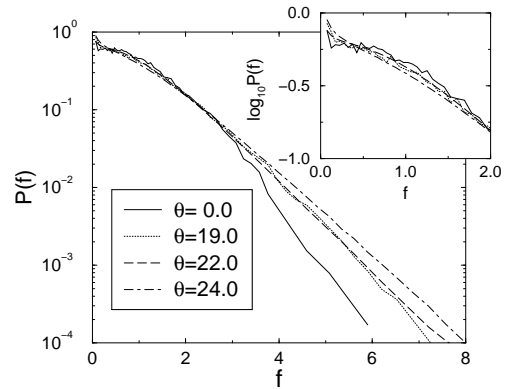} 
\bigskip
\caption{\em Distribution of normal contact forces $P(f)$ for various tilt
  angles (see text). The inset shows the emergence of a plateau in
  $P(f)$ at small forces for the static systems.}
\label{figure2}
\end{center}
\end{figure}

As shown in Fig.~\ref{figure2}, the PDFs exhibit familiar exponential
tails at high forces. When $\theta$ is reduced towards $\theta_{r}$, a
plateau develops near $f=1$ (see inset), similar to the behaviour
observed in Ref.~\cite{nagel4} near the glass transition. It may be
argued on this basis that $\theta_{r}$ appears analogous to $T_{g}$ in
a glass-former. However, the development of this feature is rather
subtle in our system, and does not appear to be a sharp indicator of
the transition from the flowing to the static case. Upon further
relaxation of the stress by reducing $\theta$ to zero, the feature
becomes more prominent similar to the results in Ref.~\cite{nagel5}.

\begin{figure}[h]
\begin{center}
\includegraphics[width=6.5cm]{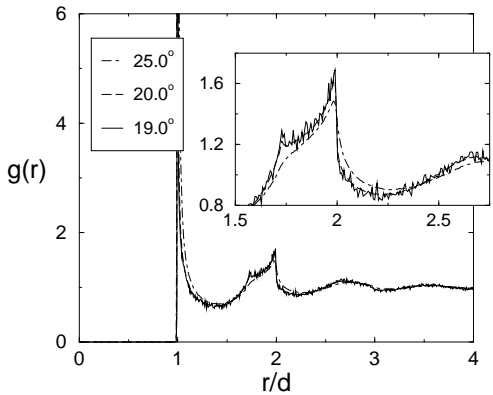}
\bigskip
\caption{\em Radial distribution function $g(r)$ for various tilt 
  angles $\theta$. The inset shows the region near the second peak.}
\label{figure3}
\end{center}
\end{figure}

As with glass transitions, we have found only subtle structural
evidence for the transition from flow to rest.  The radial
distribution function $g(r)$ is fairly insensitive to the tilt angle,
as shown in Fig.~\ref{figure3}. Although there is no long range
ordering, in the inset to Fig.~\ref{figure3} we observe the gradual
splitting of the second peak as $\theta$ is reduced. The system
develops more locally ordered structures as it becomes more compact
\cite{clarke1}.

\begin{figure}[h]
\begin{center}
  \includegraphics[width=6.5cm]{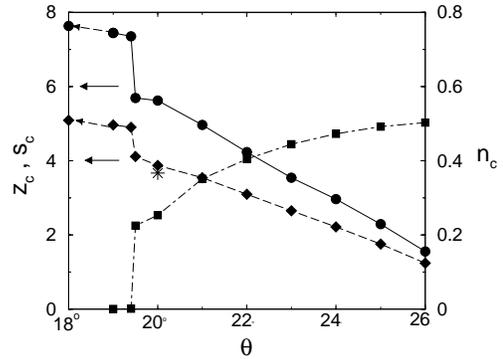} 
\bigskip
\caption{\em As the tilt angle is decreased towards $\theta_{r}$, the fraction
  of plastic contacts $n_{c}$ (squares) decreases while the
  co-ordination number $z_{c}$ (diamonds) and the staticity index
  $s_{c}$ (circles) increase. The isostatic values of $z_{c}$ and
  $s_{c}$ are indicated by the solid arrows. The dashed arrows
  indicate the values of $z_{c}$ and $s_{c}$ when we take $\theta$
  down to zero from a stopped state. The point indicated by * at
  $\theta=20^{\circ}$ is for $k_{n} = 2\cdot10^{7}mg/d$.}
\label{figure4}
\end{center}
\end{figure}

More significantly, the first peak in $g(r)$ increases with decreasing
$\theta$. Because particle neighbors are defined only on contact in
our simulations, the first peak in the radial distribution function
corresponds to the average co-ordination number $z_{c}$, which
correspondingly increases as $\theta$ decreases (see
Fig.~\ref{figure4}).

An important characteristic that separates the granular system
from a glass forming liquid is the existence of interparticle friction
and yield. To explore the influence of intergrain friction, we
distinguish between intergrain contacts at their yield criterion
$F_{t}=\mu F_{n}$ (plastic contacts), which respond plastically to
some external perturbations, from contacts with $F_{t}<\mu F_{n}$
(elastic contacts), whose response is determined by elastic
deformations of the constituent grains.  Note that the relationship
between plasticity/elasticity of individual contacts between grains
and the mechanical response of the entire system has not been explored
in this study.

\begin{figure}[h]
\begin{center}
  \includegraphics[width=6.5cm]{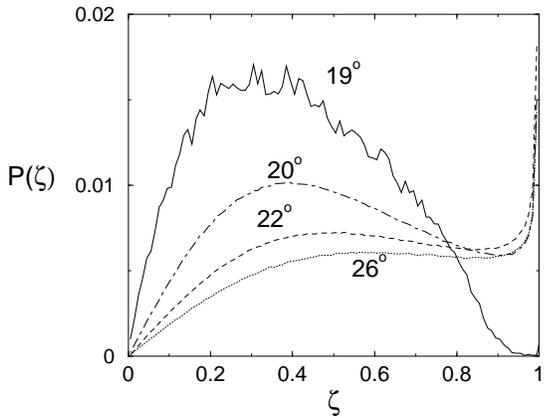} 
\bigskip
\caption{\em The probability distribution $P(\zeta)$ of frictional saturation
  $\zeta\equiv F_{t}/\mu F_{n}$ at various tilt angles. In the flowing
  state a finite fraction of plastic contacts exist (see
  Fig.~\ref{figure4}).}
\label{figure5}
\end{center}
\end{figure}

We find that the nature of grain-grain contacts is quite different in
the flowing and static regimes. The probability densities of
frictional saturations $\zeta\equiv F_{t}/\mu F_{n}$ are shown in
Fig.~\ref{figure5} for various tilt angles. The fraction $n_{c}$ of
plastic contacts (with $\zeta=1$) decreases as the tilt angle is
decreased towards $\theta_r$ (see Fig.~\ref{figure4}), accompanied
by a decrease in the average frictional saturation of elastic
contacts. For the static piles at $\theta<\theta_r$, almost all grain
contacts are elastic.

A related question is whether or not static piles of rigid grains
satisfy an isostaticity condition, where the
number of contacts is the minimum required to satisfy force and torque
balance equations for each grain \cite{moukarzel1}. This isostaticity
hypothesis has been frequently invoked in recent theoretical work
attempting to derive macroscopic constitutive relations directly from
the microstates of static granular packings\cite{grinev1}.
Isostaticity is required for a unique determination of the individual
contact forces solely in terms of the microstate.  Otherwise, the
details of grain deformations (for example, Hookean or Hertzian force
laws) must be considered in order to determine the stress state of the
pile \cite{deniz1}, as the rigid grain problem becomes ill-posed.

Counting the degrees of freedom (DOF) in the contact forces suggests
that isostaticity requires $z_{c}=4$ for frictional contacts (with 3
DOF - one vector force - per contact and 6 equations per grain) and
$z_{c}=6$ for frictionless contacts (with 1 DOF - one normal force -
per contact but only 3 equations per grain) for spherical grains.
However, one DOF is eliminated for each frictional contact that
reaches the yield criterion and becomes plastic \cite{deniz1}: the
magnitude of the tangential force in this case is determined by the
normal force and the friction coefficient $\mu$.  Thus, if the
fraction of plastic contacts is $n_c$, for a co-ordination number
$z_c$ the total number of DOF characterizing interparticle forces is
\begin{equation}
N_f=\frac{N}{2}z_c\left[3(1-n_c)+2 n_c\right],
\end{equation}
and requiring $N_f \geq 6N$ to match the number of kinematic
constraints on the particles yields
\begin{equation}
s_{c}\equiv(3-n_{c})z_{c}/2 \geq 6,
\end{equation}
where we have defined the staticity index $s_c$. The equality
corresponds to isostaticity. By analogy to the
system in Ref.\cite{deniz1}, the response of ``hyperstatic'' piles
with $s_{c}>6$ to external forces cannot be determined without
considering the history of grain contacts.
 
Figure~\ref{figure4} depicts $s_{c}$ as a function of $\theta$. As
expected, $s_{c} < 6$ for all flowing piles, though it increases with
decreasing $\theta$. However, it appears that a finite jump of $s_{c}$
at $\theta_{r}$ leaves the static pile significantly hyperstatic.
Note however that the point denoted by * in Fig.~\ref{figure4} is
for a system where $k_{n}=2\cdot10^{7}mg/d$, indicating that grain
hardness is a relevant parameter in determining $s_{c}$ and $z_{c}$
\cite{leo9}.

What do these observations teach us about the nature of the
``jamming'' transition of the granular medium?  Our earlier work
showed that granular flows obey a local rheology independent of the
history of loading \cite{deniz2,leo7}. Also, static piles are known to
generally be able to support a range of external stresses without any
macroscopic rearrangement.  So, how was the original stress state and
associated configuration of grains in this static pile determined?
One possible answer is that this state was determined by the local
loading conditions at the instant the flow ceased. This resolution is
akin to ideas discussed in the fixed principal axes (FPA)
model\cite{cates9}, except that the stress state of the static pile,
including principal stress axes, do change subsequent to flow arrest
(which is manifested as burial in the FPA model) as a function of
external loading, due to contact elasticity of grains.

We have observed analogies between jamming of granular media and the
glass transition. A plateau in the force distributions appears at
small forces once the system ceases to flow, as observed in numerical
studies of liquids at the glass transition\cite{nagel4}. Also, more
locally ordered structures appear as the system compacts. However, the
most significant effect is the transition from a flowing plastic
system with a local rheology to a static system that may preserve the
fingerprint of the stress state at the instant of jamming, which makes
this transition similar to a glass transition, in that glasses also
have an ability to be pre-stressed.  Finally, we note that we have
performed a similar study with Hookean spheres, with essentially the
same results.

Sandia is a multiprogram laboratory operated by Sandia Corporation, a
Lockheed Martin Company, for the United States Department of Energy
under Contract DE-AC04-94AL85000.  DL acknowledges support from 
US-Israel Binational Science Foundation Grant 1999235.


\begin{thebibliography}{10}

\bibitem{to1}
K. To, P.-Y. Lai, and H.~K. Pak,  Phys. Rev. Lett. {\bf 86,} 71 (2001).

\bibitem{mueth1}
D.~M. Mueth, H.~M. Jaeger, and S.~R. Nagel,  Phys. Rev. E {\bf 57,} 3164
  (1998).

\bibitem{cates2}
M.~E. Cates, J.~P. Wittmer, J.-P. Bouchaud, and P. Claudin,  Phys. Rev. Lett.
  {\bf 81,} 1841 (1998).

\bibitem{cates9}
J.~P. Wittmer, M.~E. Cates, and P. Claudin,  J. Phys. I France {\bf 7,} 39
  (1997).

\bibitem{nedderman1}
R.~M. Nedderman, {\em Statics and Kinematics of Granular Materials} (Cambridge
  University Press, Cambridge, 1992).

\bibitem{savage1}
S.~B. Savage,  J. Fluid Mech. {\bf 92,} 53 (1979).

\bibitem{liu1}
A.~J. Liu and S.~R. Nagel,  Nature {\bf 396,} 21 (1998).

\bibitem{nagel4}
C.~S. O'Hern, S.~A. Langer, A.~J. Liu, and S.~R. Nagel,  Phys. Rev. Lett. {\bf
  86,} 111 (2001).

\bibitem{nagel5}
D.~L. Blair, N.~W. Mueggenburg, A.~H. Marshall, H.~M. Jaeger, and S.~R. Nagel,
  Phys. Rev. E {\bf 63,} 041304 (2001).

\bibitem{deniz2}
D. Erta{\c s}, G.~S. Grest, T.~C. Halsey, D. Levine, and L.~E. Silbert,
  cond-mat/0005051.

\bibitem{leo7}
L.~E. Silbert, D. Erta{\c s}, G.~S. Grest, T.~C. Halsey, D. Levine, and S.~J.
  Plimpton, cond-mat/0105071.

\bibitem{pouliquen1}
O. Pouliquen,  Phys. Fluids {\bf 11,} 542 (1999).

\bibitem{footnote9}
Were we to use a sufficiently large value for interparticle friction, we would
  not expect plastic contacts even in the flowing state.

\bibitem{walton1}
O.~R. Walton and R.~L. Braun,  J. Rheo. {\bf 30,} 949 (1986).

\bibitem{cundall1}
P.~A. Cundall and O.~D.~L. Strack,  G$\acute{e}$otechnique {\bf 29,} 47 (1979).

\bibitem{footnote6}
We also take into account the rigid body motion around the contact point,
  ensuring that $\Delta{\bf s}_{t}$ always remains in the local tangent plane
  of contact.

\bibitem{footnote5}
Since the coefficient of restitution is dependent on ${\bf v}_{n}$ and goes to
  $1$ as ${\bf v}_{n}\rightarrow 0$, we increased the global damping for
  $\theta<\theta_{r}$ after the measured kinetic energy per particle
  $<10^{-3}mgd$. This was done to reduce the computer time needed to dissipate
  the remaining kinetic energy. In the static case the simulations were
  continued until the average kinetic energy per particle was less than
  $10^{-8}mgd$.

\bibitem{footgeneral}
In general, the specific values of $\theta_{r}$ and $\theta_{max}$ depend on
  $\mu$ and $\gamma_{n}$, and are sensitive to roughness of the fixed bed.

\bibitem{clarke1}
A.~S. Clarke and H. Jonsson,  Phys. Rev. E {\bf 47,} 3975 (1993).

\bibitem{moukarzel1}
C.~F. Moukarzel,  Phys. Rev. Lett. {\bf 81,} 1634 (1998).

\bibitem{grinev1}
S.~F. Edwards and D.~V. Grinev,  Physica A {\bf 263,} 545 (1999).

\bibitem{deniz1}
T.~C. Halsey and D. Erta{\c s},  Phys. Rev. Lett. {\bf 83,} 5007 (1999).

\bibitem{leo9}
L.~E. Silbert, D. Erta{\c s}, G.~S. Grest, T.~C. Halsey, and D. Levine, in
  preparation (unpublished).

\end{thebibliography}
\end{document}